\documentclass{INTERSPEECH2023}

\interspeechcameraready

\def\h{{\mathbf h}}
\def\Z{\mathcal{Z}}
\def\Y{\mathcal{Y}}
\def\H{\mathcal{H}}
\def\FSM{\mathrm{FSM}}
\def\FSync{\mathrm{FSync}}
\def\LSync{\mathrm{LSync}}

\def\CTC{\mathrm{CTC}}
\def\Enc{\mathrm{Enc}}
\def\Dec{\mathrm{Dec}}
\def\Prefix{\mathrm{Pfx}}
\def\Score{\mathrm{Score}}

\def\AncestorPruning{\mathrm{AncestorPruning}}
\def\aed{\mathrm{aed}}
\def\lm{\mathrm{lm}}
\def\fs{\mathrm{F}}
\def\ls{\mathrm{L}}
\def\fl{\mathrm{FL}}
\def\len{\mathrm{len}}
\def\ctc{\mathrm{ctc}}
\def\att{\mathrm{att}}

\def\top{\mathrm{top}}
\def\b{\mathrm{(b)}}
\def\n{\mathrm{(n)}}

\def\F{\mathcal{F}}
\def\V{\mathcal{V}}

\title{Integration of Frame- and Label-synchronous Beam Search for \\Streaming Encoder--decoder Speech Recognition}
\name{Emiru Tsunoo$^1$, Hayato Futami$^1$, Yosuke Kashiwagi$^1$, Siddhant Arora$^2$, Shinji Watanabe$^2$}
\address{
  $^1$Sony Group Corporation, Japan\\
  $^2$Carnegie Mellon University, U.S.A.}
\email{emiru.tsunoo@sony.com}

\usepackage[
backend=biber,
style=ieee,
citestyle=numeric-comp,
maxbibnames=10,
maxcitenames=3,
doi=false,isbn=false,url=false,eprint=false
]{biblatex}
\usepackage{comment}
\usepackage{algorithm}
\usepackage{algpseudocode}

\algnewcommand{\algorithmicand}{\textbf{ and }}
\algnewcommand{\algorithmicor}{\textbf{ or }}
\algnewcommand{\OR}{\algorithmicor}
\algnewcommand{\AND}{\algorithmicand}

\defbibheading{bibliography}[\refname]{}

\DeclareSourcemap{
	\maps[datatype=bibtex, overwrite=true]{
		\map{
			\step[fieldsource=booktitle,
			match=\regexp{.*Interspeech.*},
			replace={Proc. Interspeech}]
			\step[fieldsource=journal,
			match=\regexp{.*INTERSPEECH.*},
			replace={Proc. Interspeech}]
			\step[fieldsource=booktitle,
			match=\regexp{.*ICASSP.*},
			replace={Proc. ICASSP}]
			\step[fieldsource=booktitle,
			match=\regexp{.*icassp_inpress.*},
			replace={Proc. ICASSP (in press)}]
			\step[fieldsource=booktitle,
			match=\regexp{.*Acoustics,.*Speech.*and.*Signal.*Processing.*},
			replace={Proc. ICASSP}]
			\step[fieldsource=booktitle,
			match=\regexp{.*International.*Conference.*on.*Learning.*Representations.*},
			replace={Proc. ICLR}]
			\step[fieldsource=booktitle,
			match=\regexp{.*International.*Conference.*on.*Computational.*Linguistics.*},
			replace={Proc. COLING}]
			\step[fieldsource=booktitle,
			match=\regexp{.*SIGdial.*Meeting.*on.*Discourse.*and.*Dialogue.*},
			replace={Proc. SIGDIAL}]
			\step[fieldsource=booktitle,
			match=\regexp{.*International.*Conference.*on.*Machine.*Learning.*},
			replace={Proc. ICML}]
			\step[fieldsource=booktitle,
			match=\regexp{.*North.*American.*Chapter.*of.*the.*Association.*for.*Computational.*Linguistics:.*Human.*Language.*Technologies.*},
			replace={Proc. NAACL}]
			\step[fieldsource=booktitle,
			match=\regexp{.*Empirical.*Methods.*in.*Natural.*Language.*Processing.*},
			replace={Proc. EMNLP}]
			\step[fieldsource=booktitle,
			match=\regexp{.*Association.*for.*Computational.*Linguistics.*},
			replace={Proc. ACL}]
			\step[fieldsource=booktitle,
			match=\regexp{.*Automatic.*Speech.*Recognition.*and.*Understanding.*},
			replace={Proc. ASRU}]
			\step[fieldsource=booktitle,
			match=\regexp{.*Spoken.*Language.*Technology.*},
			replace={Proc. SLT}]
			\step[fieldsource=booktitle,
			match=\regexp{.*Speech.*Synthesis.*Workshop.*},
			replace={Proc. SSW}]
			\step[fieldsource=booktitle,
			match=\regexp{.*workshop.*on.*speech.*synthesis.*},
			replace={Proc. SSW}]
			\step[fieldsource=booktitle,
			match=\regexp{.*Advances.*in.*neural.*information.*processing.*},
			replace={Proc. NeurIPS}]
			\step[fieldsource=booktitle,
			match=\regexp{.*Advances.*in.*Neural.*Information.*Processing.*},
			replace={Proc. NeurIPS}]
			\step[fieldsource=booktitle,
			match=\regexp{.*Workshop.*on.* Applications.* of.* Signal.*Processing.*to.*Audio.*and.*Acoustics.*},
			replace={Proc. WASPAA}]
			\step[fieldsource=publisher,
			match=\regexp{.+},
			replace={{}}]
			\step[fieldsource=month,
			match=\regexp{.+},
			replace={{}}]
			\step[fieldsource=location,
			match=\regexp{.+},
			replace={{}}]
			\step[fieldsource=address,
			match=\regexp{.+},
			replace={{}}]
			\step[fieldsource=organization,
			match=\regexp{.+},
			replace={{}}]
		}
	}
}
\begin{document}

\maketitle
 
\begin{abstract}
Although frame-based models, such as CTC and transducers, have an affinity for streaming automatic speech recognition, their decoding uses no future knowledge, which could lead to incorrect pruning. Conversely, label-based attention encoder--decoder mitigates this issue using soft attention to the input, while it tends to overestimate labels biased towards its training domain, unlike CTC. We exploit these complementary attributes and propose to integrate the frame- and label-synchronous (F-/L-Sync) decoding alternately performed within a single beam-search scheme. F-Sync decoding leads the decoding for block-wise processing, while L-Sync decoding provides the prioritized hypotheses using look-ahead future frames within a block. We maintain the hypotheses from both decoding methods to perform effective pruning. Experiments demonstrate that the proposed search algorithm achieves lower error rates compared to the other search methods, while being robust against out-of-domain situations.
\end{abstract}
\noindent\textbf{Index Terms}: speech recognition, beam search, attention-based encoder--decoder, CTC

\section{Introduction}
Streaming style automatic speech recognition (ASR) is essential for better user experiences.
Among end-to-end ASR models, connectionist temporal classification (CTC) \cite{graves06, graves14, 
miao15, amodei16} and transducers \cite{graves13rnnt,gulati2020,zhang2020transformer} successfully model the temporal phenomenon of speech in frame-by-frame computation.
These models are referred to as frame-synchronous (F-Sync) models and they have an affinity for streaming processing \cite{dong19, yu2021fastemit, shi2021emformer}.
Conversely, the output of ASR is a label-base, which is suitable for modeling with label-synchronous (L-Sync) models, such as attention-based decoders (AttDecs) \cite{chorowski15, chan16} and language models (LMs) \cite{mikolov10, irie2019language}.
L-Sync models estimate labels in an autoregressive manner with the given context of the previously estimated output.
Generally, L-Sync models have a strong ability to model label sequences, and they are used to improve the ASR performance through fusion or rescoring \cite{chorowski17_interspeech, kannan2018analysis, sainath2019two, zhou2021phoneme}. 

During decoding, F-Sync models expand the hypotheses at each time frame.
F-Sync decoding can be efficiently computed in a beam search by maintaining scores of hypotheses ending with a blank token and non-blank tokens, respectively \cite{graves06}.
% During beam search, an LM or any L-Sync model can be incorporated into the score computation for the hypotheses ending with non-blank tokens.
However, the hypotheses are pruned at each time frame by using only partial information from the input speech.
% Although \cite{saon2020alignment} attempt to mitigate the problem, this limitation sometimes causes unreliable pruning.
This limitation sometimes causes unreliable pruning.
%In addition, the hypotheses in the beam have different lengths, for which the L-Sync model multiplies different number of probabilities; thus, they are unfair for comparison during pruning because the ranges of score are different.
Conversely, during L-Sync decoding, the hypotheses are extended token-by-token by making soft attention to all the input and the previous output sequence.
Thus, this approach has an advantage in pruning over F-Sync decoding without future information.
% , which only uses partial information.
%However, it is well-known that the decoder network is linguistically biased toward the training domain, which has recently been mitigated by estimating internal LMs \cite{meng2021ilme,,zeineldeen2021investigating}\footnote{This issue also happens in transducers.}.
%Additionally, 
However, it has a label bias problem \cite{lafferty2001conditional, bengio2015, murray2018correcting}, where once the model overestimates a label, it is difficult to downgrade it by the suffix distribution.
% Furthermore, likelihoods for stop-token degenerate for unexpectedly long outputs, which ends up with falsely generating long sequence \cite{murray2018correcting}.
Furthermore, it is difficult to use the L-Sync models in streaming ASR, where block processing is used \cite{tsunoo19, shi2021emformer} and their decoding is generally complicated \cite{moritz19, li2021head, tsunoo2022run}.
%For streaming processing, the decoder must detect the end-point in each processing blocks, which complicates the system \cite{moritz19, li2021head, tsunoo2022run}.
%Joint models have  aspect of both L-Sync and F-Sync because they are trained using the multi-task learning of the AttDec and CTC \cite{watanabe17,karita19}.
%However, the decoding is led by the AttDec, which is a L-Sync model, and CTC only predicts the scores for the top-$N$ hypotheses from the AttDec.

Both F-Sync and L-Sync decoding are complementary.
Several studies have attempted to combine F-Sync and L-Sync decoding.
Watanabe {\it et al.} proposed the joint model, where AttDec and CTC are jointly trained, and CTC rescores the hypotheses generated by AttDec during inference.
Sainath {\it et al.} proposed two-pass decoding \cite{sainath2019two}, in which the AttDec rescores $N$-best list of the transducer's hypotheses.
Li {\it et al.} combined separate F-Sync and L-Sync models where the former produces a lattice and the latter rescores it \cite{li2021combining}.
Yan {\it et al.} introduce AttDec score to F-Sync decoding of CTC in machine translation and speech translation tasks \cite{yan2022ctc}.
Additionally, \cite{dong2020comparison, zhou2021equivalence} experimentally compared F-Sync and L-Sync decoding.
However, in those studies, one is merely used for rescoring the hypotheses generated by the other beforehand, and none simultaneously considers the hypotheses from the individual decoding methods.

This study exploits complementary attributes of the F-Sync and L-Sync decoding, and proposes integrating both in a single beam search scheme.
%to mitigate the problems.
To take advantage of F-Sync decoding, which is easy to incorporate with block-wise streaming ASR, the proposed beam search primarily runs in an F-Sync manner.
%To overcome the unreliable partial F-Sync score comparison in pruning, L-Sync decoding provides the hypotheses with future knowledge that are also maintained in the beam.
%The L-Sync hypotheses have priority for survival in pruning, until all the scores of their expanded successors exceed the score of them. 
To mitigate the issue of unreliable partial F-Sync score comparison, we employ the L-Sync beam search based on the shortest token-length prefixes of the hypotheses. 
The selected hypotheses by L-Sync are then preserved in the subsequent F-Sync pruning steps and adjusted as the hypotheses expand, ensuring that the most promising ones are maintained in the proposed beam search.
Experiments demonstrate that the proposed search algorithm performs effectively pruning in the frame--label grid search, and achieve lower error rates compared to the other search methods in English and Japanese datasets.
Evaluation using various domains in each language indicates that the proposed method is robust against out-of-domain situations.

\section{F-Sync Decoding and L-Sync Decoding for Streaming ASR}
\subsection{F-Sync decoding}
\label{ssec:fsync}
%In addition to the classical hybrid neural network-hidden Markov model, 
F-Sync models include CTC \cite{graves06, graves14, 
miao15, amodei16} and transducer variants \cite{graves13rnnt,gulati2020,zhang2020transformer}, which are suitable for streaming ASR.
In streaming ASR, blockwise processing is widely used \cite{tsunoo19, shi2021emformer}.
Let $T_b$ be the last frame of $b$-th block.
An acoustic representation sequence $\H^{T_b}=\{\h_t|1\leq t \leq T_b\}$ is extracted by an encoder from speech input.
CTC and transducers introduce a blank token, $\phi$, to align $\H^{T_b}$ with a different $L$-length of label sequence $\Y^{L}=\{y_l|1\leq l \leq L\}$.
The F-Sync model estimates $\phi$-augmented tokens $z_t \in \V \cup \{\phi\}$ at time frame $t$, where $\V$ is the vocabulary.
The estimated sequence $\Z^{T'}=\{z_t|1\leq t \leq T'\}$ can be mapped to a label sequence using mapping function $\F:\Z^{T'}\rightarrow \Y^{L}$.
In the case of CTC, $T'=T_b$, and $T'=T_b+L$ for the transducers, which can emit tokens without consuming frames.
%Typically, input speech is encoded into an intermediate acoustic representation using encoder $\Enc$.
%\begin{align}
%   \h^{T} &= \Enc(\X^{T}) 
%\end{align}
Based on $\h_{t}$ and previous output label $y_{l-1}$, the F-sync model, $\FSM(\cdot)$, calculates a posterior of $z_t$.
\vspace{-0.1cm}
\begin{align}
    q_{\fs}(z_t|\h_{t},y_{l-1}) = \FSM(\h_{t}, y_{l-1})
\end{align}
For CTC, $q_{\fs}$ does not depend on the previous output.
Probability computation of $\Y^{l}$ at $t$ is efficiently performed as 
\vspace{-0.1cm}
\begin{align}
    p_{\fs}(\Y^{l}|\H^{t}) &= \gamma(\Y_\b^{l},t) + \gamma(\Y_\n^{l},t) \label{eq:fsprob} \\
    \gamma(\Y_\b^{l},t) &= \sum_{\F_{\fs}(\Z^{t})=\Y^{l}:z_t=\phi}p_{\fs}(\Y^{l}|\H^{t-1})q(z_t|\h_t,y_{l}) \nonumber \\
    \gamma(\Y_\n^{l},t) &= \sum_{\F_{\fs}(\Z^{t})=\Y^{l}:z_t=y_l}p_{\fs}(\Y^{l-1}|\H^{t-1})q(z_t|\h_t,y_{l-1}), \nonumber 
\end{align}
where $\Y_\b$ is the label sequence ending with $\phi$, and $\Y_\n$ is the other \cite{graves06}.
%by preserving the probabilities of the sequence that end with $\phi$, $\Y_\b$, and the other, $Y_\n$ \cite{graves06}.

F-Sync decoding is performed frame-by-frame.
%It is natural choice for the F-Sync models, and literature shows that the F-Sync decoding performs better than the L-Sync decoding for CTC and transducers \cite{zhou2021equivalence, yan2022ctc}.
%Although CTC and transducer have an option to use either the F-Sync decoding or the L-Sync decoding, the F-Sync decoding is a natural choice, and literature shows that the F-Sync decoding performs better than the L-Sync decoding for CTC and transducers \cite{zhou2021equivalence}.
At each time $t$, probability \eqref{eq:fsprob} is used on a logarithmic scale as F-Sync score $\alpha_{\fs}$.
\vspace{-0.2cm}
\begin{align}
    \alpha_{\fs}(\Y^{l},t) = \log p_{\fs}(\Y^{l}|\H^{t}) \label{eq:fsscore}
\end{align}
The hypotheses are copied by blank transition or expanded at each time step with $\FSync: \Y^{l} \rightarrow \Y^{l}, \Y^{l+1}$.
In the beam search with a fixed beam size $B$, the $\top{B}(\cdot)$ function returns a set of the top $B$ hypotheses at step $t$, denoted as $\Omega_{\fs,t}$:
\vspace{-0.1cm}
\begin{align}
%    \Omega_{\fs,t} = \FSync(B,\alpha(\Y,t)) = \{ \Y_{\fs}\in \top{B}(\alpha(\Y,t)) \} \label{eq:fset}
    \Omega_{\fs,t} = \top{B}(\alpha_{\fs}(\Y,t) | \Y \in \FSync(\Y))  \label{eq:fset}
\end{align}
Since the F-Sync decoding has no knowledge of future inputs even when the block processing has certain look-ahead frames \cite{tsunoo2022run, shi2021emformer}, i.e., the score is conditioned only on $\H^{t}$ at step $t$, it %more likely to 
sometimes incorrectly prune the correct hypothesis.
% equation

%The F-Sync decoding, however, is suitable for streaming ASR.
%In streaming ASR, blockwise processing is widely used \cite{tsunoo19, shi2021emformer}.
%The encoder runs in blockwise; thus the intermediate representation for $b$ blocks can be written as $\h^{T_b}$, where $T_b$ is the last frame of $b$-th block.
%Since the F-Sync requires only partial input $\H^t$ at step $t\leq T_{b}$, it can be applied directly to the streaming models without modification.

\subsection{L-Sync models with blockwise streaming processing}
\label{ssec:lsync}
We categorize AttDecs \cite{chorowski15, chan16} and LMs \cite{mikolov10, irie2019language} as L-Sync models.
At each step $i$, the L-Sync models predict the probability distribution of the next label $y_{i}$ using previous output $\Y^{i-1}$ and input $\H^{T_b}$.
L-Sync score $\beta_{\ls}$ is defined as
\vspace{-0.2cm}
\begin{align}
    \beta_{\ls}(\Y^{i}) =& \sum_{j=1}^{i} \log p_{\ls}(y_{j}|\Y^{j-1}, \H^{T_b}).  \label{eq:lsscore}
\end{align}
For LM, $p_{\ls}(y_{j})$ does not depend on input $\H^{T_b}$.
All the hypotheses are simultaneously extended with $\LSync: \Y^{i-1} \rightarrow \Y^{i}$.
Beam search at step $i$ is performed to maintain top $B$ hypotheses as in F-Sync decoding.
\vspace{-0.1cm}
\begin{align}
    %\Omega_{\ls,i} = \LSync(B,\beta(\Y^{i})) = \{ \Y_{\ls}\in \top{B}(\beta(\Y^{i})) \} \label{eq:lset}
    \Omega_{\ls,i} = \top{B}(\beta_{\ls}(\Y^{i}) | \Y^{i} \in \LSync(\Y^{i-1}))  \label{eq:lset}
\end{align}
This contextual dependency enables the L-Sync models to provide a richer representation of label sequences.
However, L-Sync models have the following two drawbacks:
\begin{itemize}
    %\item \underline{Linguistic bias problem}\footnote{Transducers have the same issue.}: L-Sync models tend to be linguistically biased toward the training domain. 
    %They do not result in sufficient performance in the mismatched domain as \cite{delrio21_interspeech} reported. 
    \item \underline{Label bias problem}: Once the model overestimates a label biased toward its training domain, the following expansion cannot easily recover it \cite{lafferty2001conditional, bengio2015, murray2018correcting}.
    \item \underline{Endpoint problem}: For streaming systems, the AttDec needs to detect an endpoint, which is the point to stop decoding with limited $\H^{T_b}$.
    This requires additional mechanisms to detect \cite{moritz19, li2021head, tsunoo2022run}.
    Once the decoding step $i$ passes the endpoint, the AttDec tends to emit unreliable tokens; thus it is better for the AttDec to proceed decoding conservatively in the streaming systems \cite{tsunoo2021slt}.
\end{itemize}

\subsection{F-Sync and L-Sync score fusion}
\label{ssec:fusion}
Because F-Sync and L-Sync models are complementary, the aforementioned problems in F-Sync and L-Sync decoding can be alleviated by fusing both scores in pruning.
\subsubsection{Score fusion in F-Sync decoding}
Typically, LMs are fused in the F-Sync decoding.
As in \cite{hannun2014deep, hwang2016character}, the scores (e.g., Eqs.~\eqref{eq:fsscore} and \eqref{eq:lsscore}) are simply combined with weights $\lambda$.
%More generally, the combined score is defined as
In the case of CTC, the combined score is defined as 
\vspace{-0.3cm}
\begin{align}
    s_{\fs}(\Y^{l}, t) = &\lambda_{\ctc} \alpha_{\ctc}(\Y^{l}, t) + \lambda_{\lm} \beta_{\lm}(\Y^{l}) \nonumber \\ &+ \lambda_{\att} \beta_{\att}(\Y^{l}) + \lambda_{\len}|\Y^{l}| \label{eq:fscore} 
%s_{\fs}(\Y^{l}, t) = &\lambda_{1} \alpha_{\fs}(\Y^{l}, t) + \lambda_{2} \beta_{\lm}(\Y^{l}) + \lambda_{3} \beta_{\att}(\Y^{l}) + \lambda_{4}|\Y^{l}| \label{eq:fscore} 
\end{align}
The last term in \eqref{eq:fscore} is a label reword, which uses the length of hypothesis $\Y^{l}$.
This mitigates unfair score comparison of different length in the beam, which, however, requires careful tuning.
%In the F-Sync beam search, two points make the decoding unstable.
%First, the first term in \eqref{eq:score} is based only on limited information of $\Z^{t}$.
%First, the first term in \eqref{eq:score} depends only on partial feature $h_t$.
%(equation)
%Second, the top $B$ hypotheses are often of different lengths.
%Thus, the L-Sync scores \eqref{eq:lsscore} are not within the same range (different $i$'s), which is not reliable for pruning.
%Although the insertion bonus mitigate the second problem, it requires careful tuning.
%We overcome these issues by proposing a new integrated beam search in Sec.~\ref{sec:integ}.

%\subsection{Streaming joint ASR model (rethink structure)}
\subsubsection{Score fusion in L-Sync decoding}
\label{ssec:joint}
%In this study, we follow the streaming encoder--decoder architecture \cite{tsunoo2022run}, which is based on the joint ASR model \cite{watanabe17}.
The joint model \cite{watanabe17} has both CTC and AttDec modules, which are trained in a multi-task learning framework.
Its decoding is based on L-Sync search led by the AttDec.
During the decoding, the score is a combination of the CTC, LM, and AttDec, as follows:
\vspace{-0.3cm}
\begin{align}
    s_{\ls}(\Y^{i},i) =& \lambda_{\ctc} \alpha_{\ctc}(\Y^{i}\dots,T_b) + \lambda_{\lm} \beta_{\lm}(\Y^{i}) \nonumber \\ 
    & + \lambda_{\att}\beta_{\att}(\Y^{i}) + \lambda_{\len}|\Y^{i}| \label{eq:jointscore} \\
    \alpha_{\ctc}(\Y^{i}\dots,T_b) =& \sum_{y_{i+1} \in \V} \alpha_{\ctc}(\Y^{i+1},T_b) \label{eq:prefix}
\end{align}
%where $\lambda_{\lm}$ and $\lambda_{\ctc}$ are tunable weight parameters.
\eqref{eq:prefix} is a CTC prefix score %\footnote{The prefix score can also use transducer scores when the transducer is jointly trained.} 
that accumulates probabilities of all the suffixes of $\Y^{i}$. % using \eqref{eq:fsscore} ($\fs=\ctc$).
The prefix scores are calculated over only the hypotheses generated by the AttDec; thus, it is sometimes difficult to recover incorrect estimation of the AttDec by itself.
Note that when $t=T_b$ and $i=l=L$, F-Sync score \eqref{eq:fsscore} and the prefix score \eqref{eq:prefix} become equivalent; thus F-Sync decoding and L-Sync decoding eventually evaluate the same score, $s_{\fs}(Y^L,T_b) = s_{\ls}(Y^L,L)$.

\section{Integrated Beam Search of the F-Sync and L-Sync Decoding}
\label{sec:integ}

%\subsection{Hypotheses combination of the F-/L-Sync decoding}
As discussed in Sec.~\ref{ssec:fsync}, the F-Sync beam search is likely to incorrectly prune the hypothesis without considering future look-ahead input, which can be prevented by the AttDec that uses all $T_b$ frames.
Conversely, the label bias problems in the AttDec (Sec.~\ref{ssec:lsync}) can be mitigated by using the F-Sync models.
Therefore, we propose maintaining both the F-Sync and L-Sync hypotheses in extended beam $B'>=B$.
Those hypotheses are generated individually from each decoding, and the step increment of $t$ of the F-Sync decoding and $i$ of the L-Sync decoding are performed alternately in an integrated single search algorithm, namely, FL-Sync beam search.

%\subsubsection{Prefix score fusion}
\subsection{Prefix score fusion}
\label{sssec:prefixfusion}
In the FL-Sync beam search, We choose F-Sync to lead decoding because it is suitable for streaming ASR.
Thus, at each step $t$, the hypotheses are copied and expanded based on F-Sync decoding as in Sec.~\ref{ssec:fsync}.
The score for hypothesis $\Y^{l}$ is extended from \eqref{eq:fscore} and \eqref{eq:jointscore} to include both $i$ and $t$ as 
\vspace{-0.2cm}
\begin{align}
 s_{\fl}(\Y^{l}, i, t) = &\lambda_{\ctc} \alpha_{\ctc}(\Y^{l}, t) + \lambda_{\lm} \beta_{\lm}(\Y^{i}) \nonumber \\&+ \lambda_{\att} \beta_{\att}(\Y^{i}) + \lambda_{\len}|\Y^{i}|. \label{eq:score} 
\end{align}
%F-Sync scores $\alpha_{\fs}$ is proved by CTC, while L-Sync scores, $\beta_{\ls}$, are provided by the AttDec and LM.
While the fist term $\alpha_{\ctc}$ defined in \eqref{eq:fsscore} is computed for $\Y^{l}$, $\beta_{\lm}$ and $\beta_{\att}$ are computed only for its prefix $\Y^{i}$, where $i \leq l$ is the current L-Sync decoding step.
The reason for using $i$ instead of $l$ is that the token prediction of the AttDec becomes unreliable when it exceeds the endpoint in streaming processing, as discussed in Sec.~\ref{ssec:lsync}.
%, the L-Sync decoding must run conservatively.
Therefore, we avoid aggressively computing the scores for the entire $l$ labels for the score fusion and conservatively use $i$ instead.
Thus, the integrated beam search aims to perform efficient pruning on this 2D grid space of $(i,t)$.

\begin{figure}[t]
%\vspace{-3.5cm}
  \centering
  %\hspace{1.5cm}
  \includegraphics[width=0.8\columnwidth]{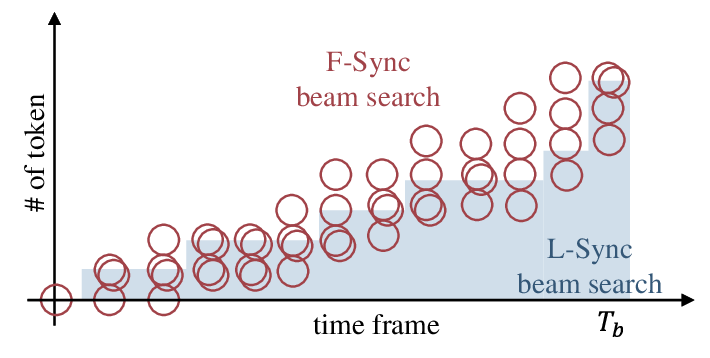}
  \vspace{-0.3cm}
  \caption{F-Sync and L-Sync decoding process.}
  \label{fig:online}
  \vspace{-0.3cm}
\end{figure}

Conversely, the L-Sync decoding expand all the $i$-length prefix of the hypotheses in hypothesis set $\Omega_{\fl,t}$, i.e., $\Y^{i} \in \Prefix_{i}(\Omega_{\fl,t})$, where $\Prefix_{i}(\cdot)$ is an $i-$length prefix extraction function.
Therefore, to perform L-Sync decoding, the minimum length of the hypotheses needs to be $\min|\Y^{l} \in \Omega_{\fl,t}| \geq i$.
The L-Sync decoding advances its step $i \rightarrow i+1$ once the minimum length becomes larger than $i$.
%The L-Sync decoding advances its step $i$ once $(i+1)$-length prefix score fusion can be performed to all the hypotheses in $\Omega_{\fl,t}$, i.e., $\min|\Y^{l} \in \Omega_{\fl,t}| = i+1$.
%Since the beam size of the L-Sync decoding is $B$, it expands $B$ prefixes of the top-$B'$ hypotheses in $\Omega_{\fl,t}$.
A graphical explanation of this is provided in Fig.~\ref{fig:online}.
The red circles indicate that the F-Sync beam search expands the hypotheses in each time step.
The L-Sync models add partial scores for the common length to these hypotheses (blue bars).
In the last step of F-Sync decoding, $T_b$, the remaining steps of L-Sync for each hypothesis are consumed, and $\beta_{\ls}(\Y^{l})$ is applied to each hypothesis to determine the best candidate.

\begin{figure}[t]
%\vspace{-3.5cm}
  \centering
  %\hspace{1.5cm}
  \includegraphics[width=0.7\columnwidth]{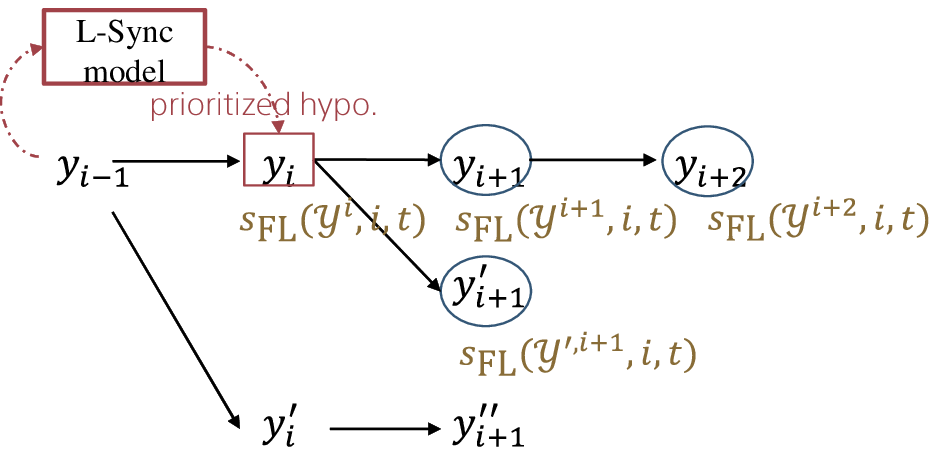}
  \vspace{-0.3cm}
  \caption{Prioritized L-Sync hypothesis (red box) and its successors (blue circles).  The score of the prioritized hypothesis is compared to all scores of the successors to perform ancestor-pruning.}
  \label{fig:decode}
  \vspace{-0.3cm}
\end{figure}

\begin{algorithm}[t]       
\hspace{-3cm}
\caption{FL-Sync beam search algorithm.}   
\label{alg:decode}                  
\footnotesize
\begin{algorithmic}[1]     
\Require last frame number of current block $T_b$, acoustic features $\H^{T_b}$, beam width for AttDec $B$, total beam width $B'$
\Ensure $\Omega_{\fl,T_b}$: hypotheses for current block $b$
\State \textbf{Initialize:} $y_0\gets\langle \mathrm{sos}\rangle$, $\Omega_{\fl,0} \gets \{y_{0}\}$, $t\gets 1$, $i\gets1$ %$\hat{\Omega} \gets \{\}$, 
\While{$t < T_b$}
%\State $\Omega_{\fl,t}\gets \{\}$ %, $\Tilde{\Omega}\gets \{\}$
\State $\hat{\Omega} \gets \{\}$
\If{$i < \min(|Y \in \Omega_{\fl,t-1}|)$}
\State $i \gets \min(|Y \in \Omega_{\fl,t-1}|)$
%\If{$m > i$}
%\State $i\gets m$
%\State $\Omega_{\fl, t} \gets \LSync(B,s_{\ls}(\Y^{i},i))$
\State $\hat{\Omega} \gets \LSync(\Prefix_{i-1}(\Omega_{\fl,t-1}))$ \Comment{Prefix L-Sync search}
\State $\hat{\Omega} \gets \top{B}(s_{\ls}(\Y^{i},i)| \Y^{i} \in \hat{\Omega})$ \Comment{Keep top-$B$ hyps}
\EndIf
%\State $\Omega_{\fl, t} \gets \{\Omega_{\fl, t}, \FSync(B'-B,s_{\fl}(\Y^{l},i,t))\}$ 
\State $\hat{\Omega} \gets \hat{\Omega} \cup \FSync(\Omega_{\fl,t-1})$  \Comment{FSync search}
%\EndIf
%\State $\Tilde{\Omega}\gets \LSync(B,s_{\ls}(\Y^{i},i))$ 
\State $\hat{\Omega} \gets \AncestorPruning(\hat{\Omega})$  \Comment{Sec. \ref{sssec:depth}}
\State $\Omega_{\fl, t} \gets$ OnlyPriority$(\hat{\Omega})$ \Comment{Sec. \ref{sssec:prefixfusion}}
\State $\Omega_{\fl, t} \gets \Omega_{\fl, t} \cup \top{(B'-|\Omega_{\fl, t}|)}(s_{\fl}(\Y,i,t)|\Y \in \hat{\Omega})$
%\State $\Theta_{\ctc} \gets$ top-$k(q_{\fs}(z_t|\H^{T_b}))$
%\For{$\Y\in\Omega_{t-1}$}
%\For{$z\in\Theta_{\ctc}$}
%\If{$z=\phi$ \OR $z=y_{|\Y|-1}$}
%\State $\hat{\Y}\gets\Y$ \Comment{copy}
%\Else
%\State $\hat{\Y}\gets\Y \circ z$ \Comment{expand}
%\EndIf
%\State $\Tilde{\Omega} \gets \{\Tilde{\Omega}, \hat{\Y}\}$
%\EndFor
%\EndFor
%\State $\Tilde{\Omega} \gets \mathrm{Merge}(\Tilde{\Omega})$
%\State $\Omega_{t} \gets$ OnlyPriority$(\Tilde{\Omega})$
%\State $\Omega_{t} \gets$ RemainingTop-$2B(\Score(\Y \in \Tilde{\Omega},t, i))$
\State $t \gets t+1$
\EndWhile
%\While {$i < |\Y_{\best}\in\Omega_{T_b}|$}
%\Comment{To determin the best hypothesis}
%\State $i \gets i+1$
%\State $\Omega_{T_b} \gets$ ReSort$(\Score(\Y \in %\Omega_{T_b},T_b,i))$
%\EndWhile
\State \Return $\Omega_{\fl,T_b}$
\end{algorithmic}
\end{algorithm}

%\subsubsection{Prioritized hypotheses from L-Sync decoding}
\subsection{Prioritized hypotheses from L-Sync decoding}
\label{sssec:priority}
The L-Sync decoding generates hypotheses $\Y_{\ls}^{i}$ based on \eqref{eq:jointscore} using all the input, including future look-ahead, which are useful to complement the weakness of F-Sync decoding.
However, the hypotheses are pruned based on \eqref{eq:score}, whose first term, $\alpha_{\ctc}(Y^{l},t)$, uses only partial $t$-frame input unlike \eqref{eq:jointscore}, and the hypotheses can still be incorrectly pruned.
To prevent from dropping L-Sync hypotheses $\Y_{\ls}^{i}$ in the early stage, we prioritize them for survival.
At step $t$, we first preserve those prioritized hypotheses of L-Sync decoding, then F-Sync decoding is performed to fill the remaining hypotheses in the beam $B'$, as
\vspace{-0.1cm}
\begin{align}
    %\Omega_{\fl, t} = &\{\Y_{\ls}^{i}\in {\top}B\left[s_{\ls}(\Y^{i},T_b)\right]\} \nonumber \\ &\cap \{\Y_{\fs}^{l}\in {\top}(B'-B)\left[s_{\fl}(\Y^{1},i,t)\right]\},
    %\Omega_{\fl, t} = &\LSync(B,s_{\ls}(\Y^{i},i)) \cup \FSync(B'-B,s_{\fl}(\Y^{l},i,t)). \label{eq:merge}
    \Omega_{\fl, t} = &\top{B}(s_{\ls}(\Y^{i},i) | \Y^{i} \in \LSync(\Prefix_{i-1}(\Omega_{\fl,t-1}))) \nonumber \\
    & \cup \top{(B'-B)}(s_{\fl}(\Y^{l},i,t) | \Y^{l} \in \FSync(\Omega_{\fl,t-1})) \label{eq:merge} 
\end{align}

\begin{comment}
In \eqref{eq:score}, we selectively use $s_{\fl}(\Y^{i}, i, T_b)$ for L-sync hypothesis $Y_{\ls}^{i}$, where prefix score \eqref{eq:prefix} is used as the F-Sync score $\alpha_{\ctc}$, as follows:
\begin{align}
    \alpha'_{\ctc}(\Y^{l},t) = \left\{ \begin{array}{ll}
    \alpha_{\ctc}(\Y^{l},t) & (\Y^{l}=\Y_{\fs}^{l}) \\
    \alpha_{\ctc}(\Y^{i} \dots ,T_b) & (\Y^{l}=\Y_{\ls}^{i}) 
    \end{array}\right.
\end{align}
The prefix score accumulates probabilities for all possible suffixes; thus $\alpha_{\ctc}(\Y^{i} \dots ,T_b) \geq \max_{t} \alpha_{\ctc}(\Y^{i},t)$ is satisfied.
By using this modified score in the pruning, $Y_{\ls}^{i}$ become most likely to survive.    
\end{comment}

\vspace{-0.3cm}
%\subsubsection{Depth-pruning for the prioritized hypothesis}
\subsection{Ancestor-pruning for the prioritized hypothesis}
\label{sssec:depth}
The priority of the hypothesis $\Y_{\ls}^{i}$ is no longer valid once the L-Sync decoding increase its step $i\rightarrow i+1$, as it generate new set of hypotheses $\Y_{\ls}^{i+1}$.
Further, to prune unnecessary hypotheses of L-Sync decoding, we propose to perform ancestor-pruning, which is similar to the depth-pruning in \cite{hwang2016character}.
As the F-Sync decoding proceed, it expands the hypothesis $\Y^{i} \rightarrow \Y^{i+1}$, as shown in Fig.~\ref{fig:decode}.
The prioritized root hypothesis $\Y^{i}$ is indicated by the red box and the expanded successors $\Y^{i+*}$ are indicated by the blue circles.
We compare the scores of all the successors and the root hypothesis, denoted as $s_{\fl}$ in Fig.~\ref{fig:decode}.
When all the scores of successors become greater than the score of the root, $s_{\fl}(\Y_{\ls}^{i},i,t) < \min(s_{\fl}(\Y^{i+*},i,t))$, we prune away prioritized root hypothesis $\Y_{\ls}^{i}$, because any new successor from the root $\Y_{\ls}^{i}$ can no longer achieve higher score than the current successors due to the $0 \leq q(z_t|\h_t,y_i) \leq 1$ restriction.

\begin{table*}[t]
  \caption{Comparison of searching algorithms of streaming ASR in English datasets in WER.  All the models were trained with Librispeech.  ID refers to in-domain. In all the decoding methods, the LM of the target domain is fused.}
  \label{tab:eng}
  %\vspace{1mm}
  \vspace{-0.3cm}
  \centering
%  \begin{tabular}{l|cc|cc}
  %\hspace{-0.5cm}
  \scalebox{0.9}{
  \begin{tabular}{l||c|c||cc|cc|c||cc}
    \hline
    & total  & L-Sync & \multicolumn{2}{c|}{LS$\rightarrow$TEDLIUM3} &  \multicolumn{2}{c|}{LS$\rightarrow$Switchboard} & LS$\rightarrow$STOP&\multicolumn{2}{c}{Librispeech (ID)} 
    \\
      & beam size& beam size & dev & test & SW & CH & & test-clean  & test-other\\
    \hline
    % No LM & & & &27.1 & 1.54\\
    Streaming L-Sync decoding \cite{tsunoo2022run}  &&5& 13.7 & 13.1 & 30.2 & 35.7 & 14.5& 3.0 & 8.1 \\
    CTC F-Sync decoding \cite{hwang2016character}  &10&& 15.0 & 14.4 & 31.0 & 37.2 & 19.2 & 3.3 & 9.0\\
    CTC+AttDec F-Sync decoding   &10&& 14.6 & 14.0& 31.4 & 36.6 &19.4 & 3.1& 8.6\\
    Integ. FL-Sync decoding (proposed)  &10&5& {\bf 13.4} & {\bf 12.7} & {\bf 29.9} & {\bf 35.0} & {\bf 14.1} & {\bf 2.9} & {\bf 7.9}\\

    \hline
  \end{tabular}
  }
  %\vspace{-0.4cm}
\end{table*}

\begin{table*}[t]
  \caption{Comparison of searching algorithms of streaming ASR in Japanese datasets in CER.  All the models were trained with CSJ+LaboroTV (LTV). ID refers to in-domain.  In all the decoding methods, the general LM is fused.}
  \label{tab:jpn}
  %\vspace{1mm}
  \vspace{-0.3cm}
  \centering
%  \begin{tabular}{l|cc|cc}
  %\hspace{-0.5cm}
  \scalebox{0.9}{
  \begin{tabular}{l||c|c||c|c|c|c||ccc}
    \hline
    & total & L-Sync & CSJ+LTV & CSJ+LTV & CSJ+LTV & CSJ+LTV &\multicolumn{3}{c}{CSJ (ID)} \\
      &beam size & beam size & $\rightarrow$TEDxJP & $\rightarrow$News& $\rightarrow$CMD1& $\rightarrow$CMD2 & eval1 & eval2 & eval3\\
    \hline
    % No LM & & & &27.1 & 1.54\\
    Streaming L-Sync decoding \cite{tsunoo2022run}  &&5& 12.1 & 3.5 & 4.1 & 12.8 & 8.0 & 7.5 & {\bf 7.0}\\
    CTC F-Sync decoding \cite{hwang2016character} &10&& 12.4 & {4.9} & 5.1& { 10.1} & 8.7&7.8&7.3\\
    CTC+AttDec F-Sync decoding  &10&& 12.1 & 3.5& 3.5& 10.7& 8.0&7.4&7.4\\
    Integ. FL-Sync decoding (proposed)  &10&5& {\bf 11.9} & {\bf 3.0} & {\bf 3.2} & {\bf 9.6} & {\bf 7.9} & {\bf 7.3} & 7.1 \\
    
    \hline
  \end{tabular}
  }
  %\vspace{-0.4cm}
\end{table*}
\subsection{FL-Sync beam search algorithm}
The proposed beam search is summarized in Algorithm~\ref{alg:decode}. %, and it is graphically shown in Fig.~\ref{fig:decode}.
As the FL-Sync decoding is performed based on the F-Sync frame-wise pruning, $t$ is increased at each step in the loop.
First, L-Sync decoding is performed if step $i<\min(|Y \in \Omega_{\fl,t-1}|)$ as descried in Sec.~\ref{sssec:prefixfusion}.
The L-Sync decoding expands the prefixes of the hypotheses (Line~6) and provides prioritized hypotheses $\Y^i_{\ls}$, which are maintained in temporary hypothesis set $\hat{\Omega}$ (Line~7).
Then the F-Sync decoding is performed based on \eqref{eq:score}, and merged with the prioritized hypotheses (Line~9). %, which corresponds to \eqref{eq:merge}.
Subsequently, in Line~10, the ancestor-pruning described in Sec.~\ref{sssec:depth} is performed, which removes unnecessary hypotheses.
The hypotheses from L-Sync decoding have priority for survival (Line~11), then lastly the hypothesis set is filled from the temporary set up to total beam size, $B'$ (Line~12).
Note that, even the total beam size increases from $B$ to $B'$, computational impact is marginal from $B$-beam L-Sync decoding, because the F-Sync model, e.g., CTC, generally requires much less computation.

\section{Experiments}
To evaluate the effectiveness of the proposed FL-Sync beam search, we evaluated cross-domain and intra-domain scenarios in English and Japanese datasets.

%\subsection{Search algorithm comparison in English datasets}
%\label{ssec:english}
%\subsubsection{Experimental setup}
\subsection{Experimental setup}
\label{ssec:english_setup}
For the English evaluation, we trained a streaming encoder--decoder ASR model following \cite{tsunoo2022run} with the Librispeech dataset \cite{panayotov15}, a read speech corpus.
It was applied to three various domains: The TED-LIUM 3 \cite{hernandez2018ted} dev/test set (a spontaneous lecture style), Hub5’00 (telephony-style conversation) having Switchboard (SWB) and CallHome (CHM) subsets, and voice-command-style STOP dataset \cite{tomasello2023stop}.

For Japanese, we trained a streaming ASR model with a merged set of the lecture style CSJ \cite{csj} and LaboroTV corpus of TV program \cite{ando2021construction}.
We used three evaluation sets of the CSJ dataset for intra-domain setup, and TEDxJP \cite{ando2021construction} for cross-domain setup.
To cover various range of domains, We also used in-house evaluation data; 720 read news utterances (News) spoken by four males and four females, 2,620 voice commands (CMD1) and far-field 1,768 commands (CMD2), each spoken by 10 hired speakers.

The input acoustic features were 80-dimensional filter bank features.
The input features were applied mean normalization with a sliding window.
The encoder consisted of 12 blocks of conformer \cite{gulati2020}.
The AttDec had six decoder blocks.
Each block consisted of four-head 256-unit attention layers and 2048-unit feed-forward layers.
Contextual block encoding \cite{tsunoo19} was applied to the encoder with a block size of 40, a shift size of 16, and a look-ahead size of 16.
The models were trained using multitask learning with CTC loss \cite{watanabe17}, with a weight of 0.3.
%We used the Adam optimizer and Noam learning rate decay, and applied SpecAugment \cite{park19}.

For English evaluation, external LMs were trained using each target dataset for adaptation.
LMs were four-layer unidirectional LSTM with 2048 units, using the byte-pair encoding (BPE) subword tokenization with 5000 token classes.
%The LMs were fused with a weight of 0.6/0.4 for the cross-/intra-domain evaluation in English ($\lambda_{\lm}$ in \eqref{eq:score}).
For Japanese, we collected over 27 million sentences to train a general LM with 10,000 BPE tokens, applied to all the tasks.

We compared our proposed FL-Sync decoding with streaming L-Sync decoding \cite{tsunoo2022run}, and CTC F-Sync decoding \cite{hwang2016character}.
We also evaluated AttDec score fusion for CTC F-Sync decoding for comparison.
%For LM fusion in CTC decoding, we set the LM weight ($\lambda_{\lm}$ in \eqref{eq:fscore}) as 0.4, and did not apply the AttDec score ($\lambda_{\att}=0$).
%For the CTC F-Sync decoding methods, $(\lambda_{\lm}, \lambda_{\att})=(0.4,0)$ or $(0.4,0.4)$ were used for English datasets, and $\lambda_{\lm}=0.1$ was used for Japanese, instead.
For the CTC F-Sync decoding methods, $\lambda_{\lm}=0.4$ was used for English datasets, and $\lambda_{\lm}=0.1$ for Japanese, instead.
For the baseline L-Sync decoding and the proposed FL-Sync decoding, we adopted $(\lambda_{\lm}, \lambda_{\ctc})=(0.4,0.4)$ for English intra-domain, $(0.6,0.4)$ for English cross-domain, and $(0.3,0.5)$ for Japanese experiments.
$\lambda_{\att}$ was set as $(1-\lambda_{\ctc})$.
We adopted $\lambda_{\len}=1$ for all the evaluation.
We used $B'=10$ and $B=5$ as a beam size for total FL-Sync decoding and L-Sync decoding in it, respectively.
For fair comparison, we used the same beam sizes for the baseline approaches.
%In the baseline L-Sync, the hypotheses were expanded to $B\times 3B$ to calculate CTC prefix score $\alpha_{\ctc}$ in \eqref{eq:attscore} based on the top $3B$ hypotheses of the AttDec.
%For CTC F-Sync decoding, the hypotheses were expanded to $B \times B$.
%In the proposed FL-Sync, it followed F-Sync expansion with a total beam size of $2B$; thus the hypotheses were expanded to $2B \times B$, which was smaller than the baseline L-Sync space.

\subsection{Results of English evaluation}
The word error rate (WER) results are summarized in Table~\ref{tab:eng}.
%Streaming L-Sync decoding was generally better than CTC F-Sync decoding, which indicated that AttDec had more ability to model sequence, and were effective to be fused with LMs.
Streaming L-Sync decoding was generally better than CTC F-Sync decoding, particularly in the voice-command STOP dataset.
This indicated that it is more effective in pruning to use input information including future look-ahead frames.
Although the AttDec score fusion improved accuracy on CTC F-Sync decoding, the proposed FL-Sync decoding showed large improvement from those F-Sync decoding methods because it also preserved hypotheses from the L-Sync decoding.
When we compare FL-Sync decoding with the L-Sync baseline decoding, our proposed method performed robustly in the cross-domain scenarios; for instance, WER was improved from 13.1\% to 12.7\% in TEDLIUM3 test set.
Furthermore, we confirmed that it did not degrade the intra-domain performance, or even observed slight improvement from 8.1\% to 7.9\% in test-other, for instance.
Since the scores, \eqref{eq:fscore}, \eqref{eq:jointscore}, and \eqref{eq:score}, eventually become the same in all the search methods, the results show that our method is effective in pruning because of maintaining hypotheses from both L-Sync and F-Sync decoding.
%Thus, FL-Sync ended with better results than the baseline L-Sync decoding.

\subsection{Results of Japanese evaluation}
We evaluated character error rates (CERs), which are listed in Table~\ref{tab:jpn}.
The results followed similar tendency to the English evaluation.
As we used the general LM, it did not exactly match to some of the target domains.
%As the result, we observed that CTC F-Sync decoding had difficulty in fusing LM.
As a result, we observed that even in the source domain CSJ eval sets, CERs were relatively higher than the other literature \cite{tsunoo2022run, karita2021comparative}. 
Furthermore, the baseline L-Sync decoding had difficulty in adapting to the target domains, in particular in CMD2 data, which tend not to be grammatically structured.
Since our proposed FL-Sync decoding maintained both L-Sync and F-Sync decoding, it recovered the domain bias and was robust to the cross-domain situation.

\section{Conclusion}
We have proposed a new FL-Sync beam search, which integrates complementary F-Sync and Sync decoding to mitigate the problems in both the decoding.
The proposed beam search primarily runs in an F-Sync manner to incorporate it with block-wise streaming ASR.
To overcome the unreliable partial F-Sync score comparison in pruning, L-Sync decoding provide the prioritized hypotheses with future look-ahead input frames, which are also maintained in the beam to perform effective pruning.
%The L-Sync hypotheses have priority for survival in pruning, until all the scores of expanded successors exceed the score of them. 
%Experiments showed that the proposed search algorithm had an advantage in performance comparing to the baseline F-Sync and L-Sync search.
%We also observed that FL-Sync also had more robustness against out-of-domain situations than L-Sync decoding.

%\bibliographystyle{IEEEtran}
%\bibliography{mybib}
\section{References}
{
\printbibliography
}
\end{document}